\documentstyle[aps,prb,epsfig]{revtex}
\begin{document}
\draft
\author{Tomosuke Aono$^1$ and
Mikio Eto$^{2}$}
\address{$^1$The Institute of Physical and Chemical Research (RIKEN),\\
2-1 Hirosawa, Wako-shi, Saitama 351-0198 Japan\\
$^2$Faculty of Science and Technology, Keio
University,\\ 3-14-1 Hiyoshi,
Kohoku-ku, Yokohama 223-8522 Japan}
\title{Kondo effect in coupled quantum dots under magnetic fields}
\date{\today}
\maketitle
\begin{abstract}
The Kondo effect in coupled quantum dots is investigated theoretically
under magnetic fields.
We show that the magnetoconductance (MC) illustrates
peak structures of the Kondo resonant spectra.
When the dot-dot tunneling coupling $V_C$ is smaller than
the dot-lead coupling $\Delta$ (level broadening),
the Kondo resonant levels appear at the Fermi level ($E_F$).
The Zeeman splitting of the levels
weakens the Kondo effect, which results in a negative MC.
When $V_{C}$ is larger than $\Delta$, the Kondo resonances form bonding
and anti-bonding levels,
located below and above $E_F$, respectively.
We observe a positive MC since the Zeeman splitting increases the
overlap between the levels at $E_F$.
In the presence of the antiferromagnetic spin coupling between the dots,
the sign of MC can change as a function of the gate voltage.
\end{abstract}

The microfabrication technique on semiconductors has enabled us to
make quantum dots with the size of the order of the Fermi wavelength.
Such dots are often referred to as ``artificial atoms" due to the
discreteness of their energy levels.\cite{Tarucha,METbook}
By coupling the artificial atoms,
we can fabricate ``artificial molecules." Indeed,
interdot ``molecular orbitals'' have been observed experimentally
when the tunneling coupling between the dots $V_C$ is sufficiently
large.\cite{Oosterkamp,Blick}
The strength of $V_C$ can be controlled by external gate voltages.

Recently the Kondo effect has been found in single quantum dot
systems.\cite{Goldhaber,Cronenwett,Simmel99,Schmid00,Wilfred00}
The dot-lead coupling plays a role in the Kondo effect,
the strength of which is characterized by the level
broadening $\Delta = \pi \rho V^2$ where $\rho$ is the density of
states in the leads and $V$ is the tunneling probability amplitude
between the dot and leads.
When a localized spin in a dot is coupled
to the Fermi sea in the leads, the dot spin is screened out and a
resonant level is formed at the Fermi level $E_F$.
The resonant width is of the order of the Kondo temperature $T_K$.
The conduction electrons can be transported through the Kondo
resonant level, which results in the unitary limit of
the conductance,
$G=2e^2/h$.\cite{Glazman,Ng,Kawabata,Hershfield,Meir93,Wingreen94}

In coupled quantum dots connected in series,
various Kondo phenomena have been proposed
theoretically.\cite{Ivanov,Pohjola,Aono98,Georges,Andrei,Busser99,Aguado99,Izumida00,Aono01}
In our previous papers,\cite{Aono98,Aono01}
we have pointed out the importance of competition between the
dot-dot coupling $V_C$ and dot-lead coupling $\Delta$.
When the dot-lead coupling larger than the dot-dot coupling
($V_C<\Delta$), the Kondo resonances are created between a dot and a lead.
The Kondo resonant levels appear at $E_F$ and the electron transport is
determined by the hopping probability between the resonant levels.
When $V_C>\Delta$, the Kondo levels are split into two, forming
bonding and anti-bonding levels. They are located below and above
the Fermi level, $E_F \mp T_K \sqrt{({V}_{\rm C}/\Delta)^2-1}$,
and consequently the conductance is suppressed.

These unique characters of the Kondo resonant spectra in coupled
dots can be observed directly. Aguado and Langreth have calculated the
differential conductance under finite source-drain voltages
$V_{sd}$.\cite{Aguado99}
They have shown that the $dI/dV_{sd}$ curve has a single peak at $V_{sd}=0$
when $V_C<\Delta$ and it has double peaks when
$V_C > \Delta$.
In this letter, we propose an alternative probe for the resonant
spectra, a magnetoconductance (MC).
In single quantum dot systems,
the Zeeman effect lifts off the degeneracy of the spin states and
hence weakens the Kondo effect. In consequence the conductance $G$
decreases with increasing magnetic field (negative MC).\cite{Meir93}
In coupled dots, the situation is the same when $V_C<\Delta$.
The Kondo resonant levels are split into two by the Zeeman effect,
below and above $E_F$ for spin down and up electrons, respectively,
which reduces $G$.
For $V_C>\Delta$, on the other hand, we observe a positive MC.
The Zeeman splittings increase the overlap between the bonding resonant level
for spin up electrons and anti-bonding resonant level for spin down electrons
at $E_F$. This enhances the conductance $G$. When the Zeeman energy exceeds
$T_K$, the Kondo effect is broken and $G$ drops to zero suddenly.

In the above discussion, we have disregarded the antiferromagnetic
spin coupling $J$ between the dots.
Since the dot-lead Kondo coupling
screens the dot spin,
the dot-dot spin coupling $J$ competes with the
Kondo coupling.\cite{Jones,Jones2,Sakai90}
Georges and Meir have illustrated
various transport properties in $J$ vs.\ $V_C/\Delta$ plane.\cite{Georges}
We have shown that the effect of $J$ on the
conductance $G$ can also be understood in terms of the resonant
spectrum.\cite{Aono01} In the second half of this letter,
we demonstrate the calculations of MC in the presence of $J$.
We obtain an interesting result when $V_C/\Delta < 1$:
The sign of MC changes as a function the gate voltage $V_g$.
For small negative $V_g$, the Kondo coupling is larger than the spin-spin
coupling. We find a negative MC.
For large negative $V_g$, the spin coupling $J$ increases
the effective dot-dot coupling, which results in a positive MC.

\begin{figure}[tbp]
    \centering
\includegraphics[width=0.5\textwidth]{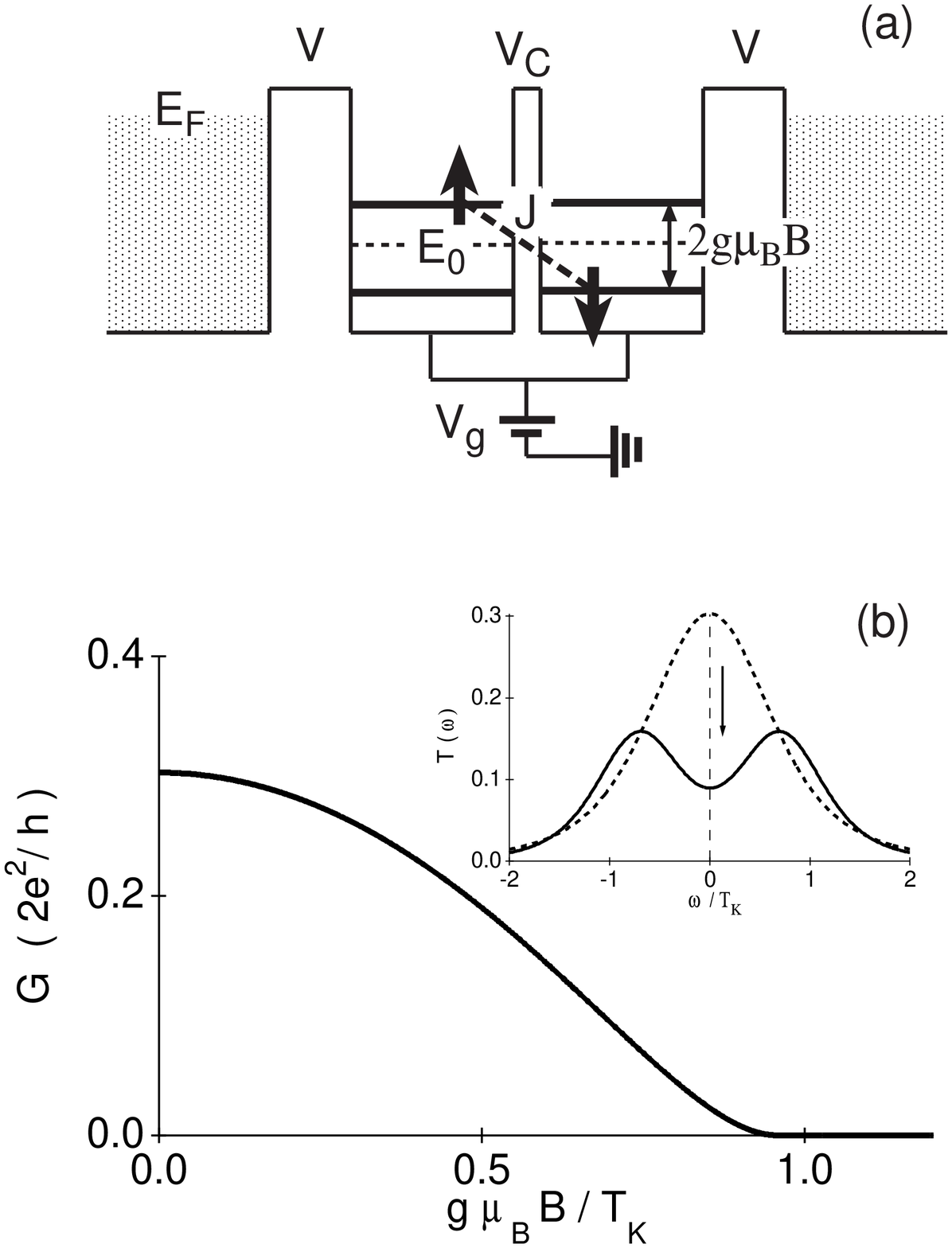}
    \caption{%
(a) Coupled quantum dots in the presence of the Zeeman
    effect, $ g \mu_B B$, under a magnetic filed $B$.
    The antiferromagnetic spin-spin coupling is denoted by $J$.
(b) Conductance $G$ as a function of the Zeeman splitting $g \mu_B B$
when $V_C / \Delta < 1$ ($V_C / \Delta = 0.3$), in the absence of the
antiferromagnetic spin coupling $J$. $E_0 / \Delta = -2$.
$g \mu_B B$ is normalized by the Kondo temperature $T_K$ at
$B=0$.
Inset: Transmission probability $T(\omega)$ 
with $\omega$ being the energy of an incident electron.
The dotted and solid lines represent the cases of $B=0$ and
$g \mu_B B / T_K = 0.71$, respectively.
The arrow indicates the direction of an increase in magnetic
field $B$.}
    \label{fig:MC_weak}
\end{figure}

As a model,
we consider a symmetric quantum dot dimer connected in series
as shown in Fig.\ \ref{fig:MC_weak}(a).
Each dot has a single energy level $E_0$ and accommodates an electron
with spin up or down.
The energy level is split into $E_0 \pm g \mu_B B$
by the Zeeman effect under a magnetic field $B$.
A common gate voltage $V_g$ is attached to the dots to control $E_0$.
Two dots couple to each other with $V_{C}$,
and to external leads with $V$.
We assume that the intradot Coulomb interaction $U$ is
sufficiently large so that
(i) the double occupancy of electrons in each dot is forbidden, but
(ii) the antiferromagnetic spin coupling exists
between the quantum dots, through the virtual double occupancy
in a dot: $J {\bbox S}_{L} \cdot {\bbox S}_{R}$ where $J=4 V_{C}^2/U$
and ${\bbox S}_{\alpha}$ is the spin operator in dot $\alpha=L,R$.
The interdot Coulomb interaction is neglected.

Let us disregard the antiferromagnetic spin coupling $J$ for a while.
The Hamiltonian reads
\begin{eqnarray}
 {\cal H}_0 &=&
 \sum_{\stackrel{\alpha={\rm L, R}}{k, \; \sigma = \uparrow,\downarrow}}
 E(k) c^{\dag}_{\alpha k \sigma}c_{\alpha k \sigma}
+
\sum_{\stackrel{\alpha={\rm L, R}}{\sigma} }
\left( E_0 + \sigma g \mu_B B \right) C^{\dag}_{\alpha \sigma} C_{\alpha
\sigma} \nonumber \\
&+& \frac{V}{\sqrt{2}}  
\sum_{\alpha, k ,\sigma } \left(
c^{\dag}_{\alpha k \sigma} C_{\alpha \sigma} + {\rm H.c.} \right)
+\frac{V_{C}}{2} 
\sum_{\sigma} \left(
C^{\dag}_{L \sigma} C_{R \sigma} +
{\rm H.c.} \right),
\label{eq:H_orig}
\end{eqnarray}
where
$c^{\dag}_{\alpha k \sigma}$ creates an electron in
lead  $\alpha = L, R$
with
energy $E(k)$ and spin $\sigma$,
and $C^{\dag}_{\alpha \sigma}$ creates an electron in
dot $\alpha$ with spin $\sigma$.
The prohibition of double occupancy in each dot is required.
To treat this situation, we adopt the slave boson
formalism.\cite{Coleman87,Read83,Bickers87,Newns88,HewsonBook}
The annihilation operator of an electron in dot $\alpha$
is decomposed as $C_{\alpha \sigma}=b^{\dag}_{\alpha} f_{\alpha \sigma}$,
where a slave boson operator $b^{\dag}_{\alpha}$
creates an empty state and a fermion operator $f_{\alpha \sigma}$
annihilates a singly occupied state with spin $\sigma$.
The merit of this formalism is that
the prohibition of double occupancy is expressed by equations:
$
Q_{\alpha} \equiv \sum_{\sigma} f^{\dag}_{\alpha \sigma} f_{\alpha \sigma} +
b^{\dag}_{\alpha} b_{\alpha}=1.
$
These constraints can be taken into account by adding a term,
$ \sum_{\alpha} \lambda_{\alpha} \left( Q_{\alpha} -1 \right)$,
to the Hamiltonian (\ref{eq:H_orig})
with the Lagrange multipliers $\lambda_{\alpha}$.

We apply a mean field theory for the Kondo effect to our
model.\cite{Georges,Jones2,Coleman87}
It is a semi-quantitative theory and useful to elucidate the Kondo
resonant spectra.
The operator $b_{\alpha}(t)$ is replaced by a constant real number,
$b_{\alpha}(t) = b_{\alpha}$. By the symmetry of the system,
$b_L=b_R \equiv b$ and $\lambda_L=\lambda_R \equiv \lambda$.
Then the problem is reduced to a one-body problem,\cite{Kawamura}
\begin{eqnarray}
{\cal H} &=&
\sum_{\stackrel{\alpha={\rm L, R}}{k, \; \sigma}}
 E(k) c^{\dag}_{\alpha k \sigma}c_{\alpha k \sigma}
+
\sum_{\stackrel{\alpha={\rm L, R}}{\sigma}}
\left( \widetilde{E}+\sigma g \mu_B B \right) 
f^{\dag}_{{\alpha} \sigma} f_{{\alpha} \sigma} \nonumber \\
&+&
\widetilde{V}_C \sum_{\sigma} \left( f^{\dag}_{{L} \sigma}
f_{{R} \sigma} + {\rm H.c.} \right)
+
\widetilde{V}
\sum_{\alpha, k, \sigma}
\left( c^{\dag}_{\alpha k \sigma}f_{\alpha \sigma}  +
{\rm H.c.} \right) \nonumber\\
&+& \lambda \sum_{\alpha = {\rm L, R}} ( b^{2} - 1 ),
\label{eq:HMF}
\end{eqnarray}
with an ``energy level'' $\widetilde{E}=E_0+\lambda$ in the dots and
``tunneling couplings'' $\widetilde{V}_{C} = b^2 V_{C}/2$,
$\widetilde{V}= b V/\sqrt{2}$.
We determine $b$ and $\lambda$ by minimizing the expectation value
of the Hamiltonian (\ref{eq:HMF}). Then the effective dot-lead coupling,
$\widetilde{\Delta}=\pi \rho \widetilde{V}^2=b^2\Delta/2$, is equal to
the Kondo temperature $T_K$.\cite{Kondo_temp}
Note that $\widetilde{V}_{C}/\widetilde{\Delta} =
V_C /\Delta$.\cite{Aono98,Aono01}
The linear-conductance $G$ is written as
\begin{equation}
G = \frac{e^2}{h}
 \sum_{\sigma = \uparrow, \downarrow} T_{\sigma} (\omega = 0)
 \equiv \frac{2 e^2}{h} T(\omega=0)
\label{Eq}
\end{equation}
with the transmission probability
$T_{\sigma}(\omega)$ for an incident electron with spin $\sigma$ and
energy $\omega$.\cite{Aono98}
We have chosen $\omega=0$ at the Fermi level in the leads.

First, we study the case of $V_C / \Delta < 1$
($\widetilde{V}_{C}/\widetilde{\Delta}<1$).
In Fig.\ \ref{fig:MC_weak}(b),
the conductance $G$ is plotted as
a function of $B$ when
$V_C/\Delta = 0.3$.
As $B$ increases,
$G$ decreases monotonically
(negative MC).
At $ g \mu_B B \simeq T_{K}$, the Kondo coupling disappears and $G=0$,
where electrons in the dots are isolated from the leads and
make a spin polarized state by the Zeeman effect.
The inset in Fig.\ \ref{fig:MC_weak}(b) shows 
the transmission probability $T(\omega)$.
When $B=0$, $T(\omega)$ has a single peak at $\omega =0$
(dotted line). This is because the Kondo resonant levels are formed
between a dot and a lead at $\omega =0$. The electron transport is
determined by the hopping between the resonant states,
and hence the peak height of $T(\omega)$ is less than unity.
When $B \neq 0$, the Zeeman effect
splits the Kondo levels by $g \mu_B B$. In consequence
$T(\omega)$ has double peaks at $ \omega \simeq \pm g \mu_B B$
(solid line).
With increasing $B$, the peaks are separated more, which
results in an decrease in $T(\omega =0)$.

\begin{figure}[tbp]
    \centering
\includegraphics[width=0.5\textwidth]{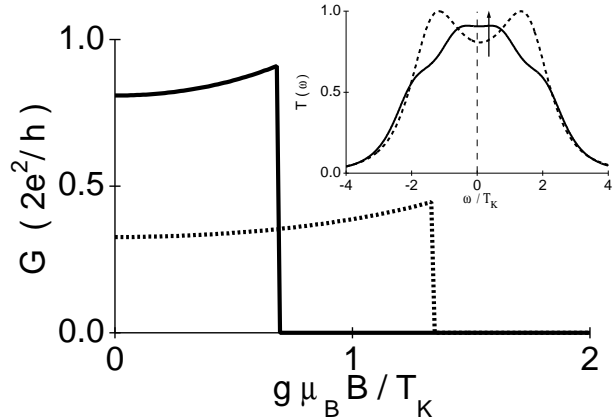}
    \caption{%
Conductance $G$ as a function of the Zeeman splitting $g \mu_B B$
when $V_C / \Delta >1$ ($V_C / \Delta = 1.6$).
The solid and dotted lines represent the cases of $J=0$ and
$J/\Delta=1.0$, respectively. $E_0 / \Delta = -1.5$.
$g \mu_B B$ is normalized by the Kondo temperature $T_K$ at $B=0$.
Inset:
Transmission probability $T(\omega)$
with $\omega$ being the energy of an incident electron,
in the absence of $J$. 
The dotted and solid lines represent the cases of $B=0$ and
$g \mu_B B / T_K = 0.63$, respectively.
The arrow indicates the direction of an increase in magnetic
field $B$.}
    \label{fig:MC_strong}
\end{figure}

Next, we study the case of $V_C / \Delta > 1$
($\widetilde{V}_{C}/\widetilde{\Delta}>1$).
In Fig.\ \ref{fig:MC_strong}, $G$ is plotted as a function of $B$
when $V_C/\Delta = 1.6$ (solid line).
As $B$ increases from zero, $G$ increases (positive MC).
This result is in contrast to that in the case of $V_C / \Delta < 1$.
At $ g \mu_B B \sim T_{K}$, the Zeeman effect destroys
the Kondo coupling and $G$ drops to zero suddenly.
The inset in Fig.\ \ref{fig:MC_strong} presents $T(\omega)$.
When $B=0$, $T(\omega)$ has double peaks at
$\omega =\mp \sqrt{\widetilde{V}^2_{\rm C}-\widetilde{\Delta}^2}$
(dotted line).
These peaks correspond to the molecular levels between the Kondo
states.\cite{Aono98} When $B \neq 0$, the Zeeman effect
splits both of these molecular levels. There are totally four peaks
(solid line).
As $B$ increases, the middle two peaks are overlapped more at $E_F$,
which leads to an increase in $T(\omega = 0)$.

Now we consider the antiferromagnetic spin coupling between the dots,
$J {\bbox S}_{L} \cdot  {\bbox S}_{R} =
(J/2) \sum_{\sigma,\sigma'} \;
(f^{\dag}_{{R} \sigma}  f_{{L} \sigma}
f^{\dag}_{{L} \sigma'}  f_{{R} \sigma'}
+ {\rm H.c.})$.
We introduce an order parameter for the spin-spin coupling,
$\kappa = (J/2) \sum_{\sigma} \langle
f^{\dag}_{{R} \sigma} f_{{L} \sigma} \rangle$,
to decouple the spin-spin interaction\cite{Jones2,Coleman87}
\begin{equation}
J {\bbox S}_{L} \cdot  {\bbox S}_{R} \rightarrow 
\frac{\kappa^2}{J} +
 \sum_m \kappa \left( f^{\dag}_{L m} f_{R m} + {\rm H.c.} \right).
\end{equation}
Thus $\widetilde{V}_C$ in Eq.\ (\ref{eq:HMF}) is replaced by
$\kappa + \widetilde{V}_C$.\cite{Aono01}
The spin-spin coupling $\kappa$, therefore, increases the dot-dot
coupling effectively.
$T(\omega)$ has a single peak at $\omega=0$ when
$\kappa + \widetilde{V}_C < \widetilde{\Delta}$, and double peaks
below and above $\omega=0$ when
$\kappa + \widetilde{V}_C > \widetilde{\Delta}$.

\begin{figure}[tbp]
    \centering
 \includegraphics[width=0.5\textwidth]{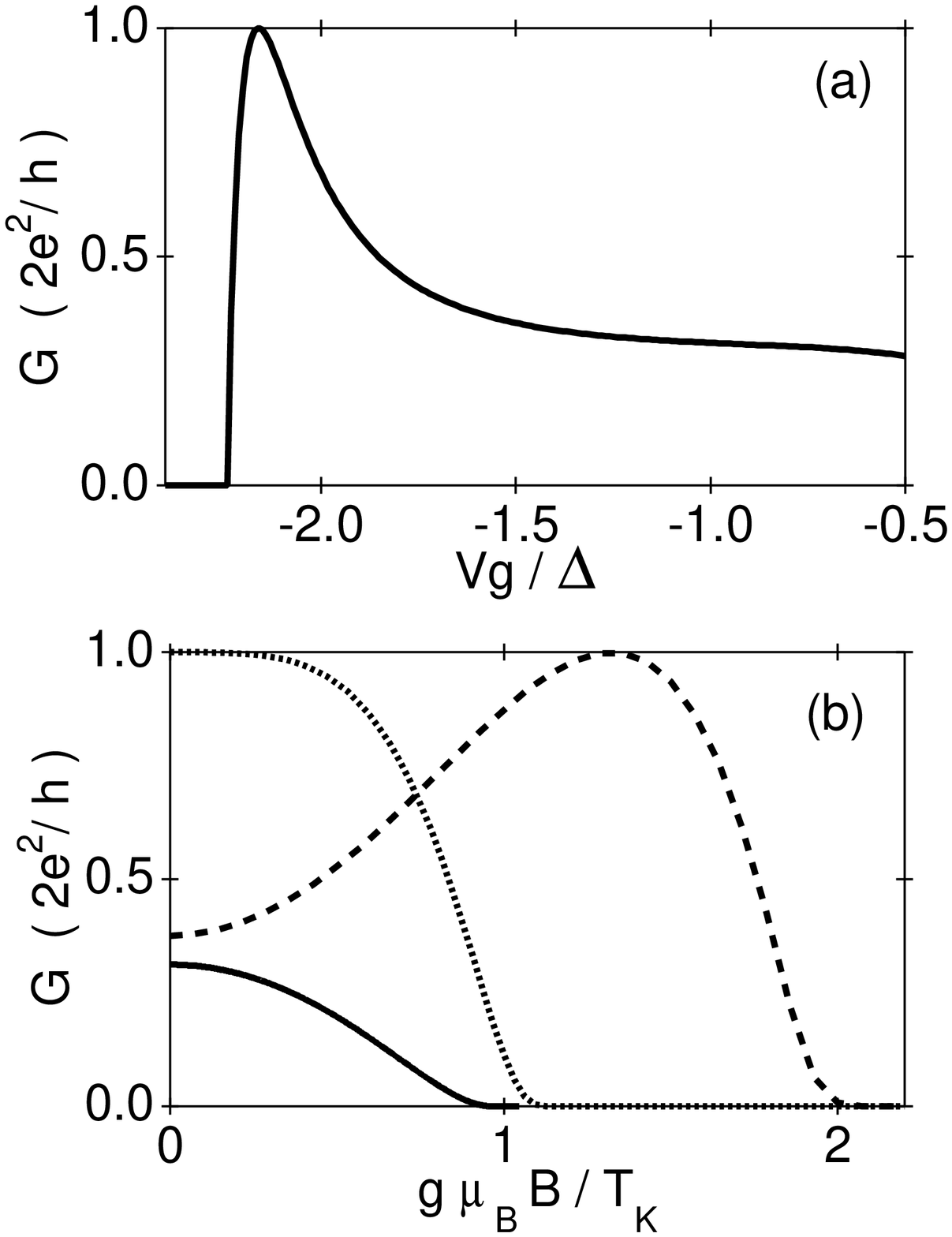}   
    \caption{(a) The gate voltage dependence of the conductance $G$
    when $V_C/ \Delta = 0.3$ and $J / \Delta = 9.0 \times10^{-4}$.
    $B=0$. We define $V_g=E_0$.
(b) Conductance $G$ as a function of the Zeeman splitting
$g \mu_B B$. $V_g / \Delta = -1.0$ (solid line), $-2.16$ (dotted line),
and $-2.23$ (broken line).
$g \mu_B B$ is normalized by the Kondo temperature $T_K$ at $B=0$.}
    \label{fig:G_MC_J_weak}
\end{figure}

Let us discuss the case of $V_{C} / \Delta < 1$.
In Fig.\ \ref{fig:G_MC_J_weak}(a),
$G$ is plotted as a function of the gate voltage $V_g$ when $B=0$.
$V_C/\Delta = 0.3$ and $J/\Delta = 9.0 \times 10^{-4}$.\cite{J}
The gate voltage changes the dot level $E_0$ (we define $V_g=E_0$).
As $V_g$ is smaller, the Kondo coupling $\widetilde{\Delta}$ $(=T_K)$
is weaker.\cite{Kondo_temp}
(i) At sufficiently large negative $V_g$,
$\kappa = J/2$ and $\widetilde{\Delta} = 0$.
This indicates that a spin-singlet state appears between the dot spins
and the Kondo coupling between a spin and lead completely disappears.
The conductance $G=0$. With increasing $V_g$,
$\widetilde{\Delta}$ begins to increase
whereas $\kappa$ becomes smaller.
(ii) At sufficiently small negative $V_g$,
the electronic state is dominated
by the Kondo coupling and spin-spin coupling is ineffective
($\kappa \ll T_K$).
Then $\kappa + \widetilde{V}_C<\widetilde{\Delta}$ and
the transmission probability $T(\omega)$ has a single peak at
$\omega=0$. The peak height is less than unity and hence
$G$ takes a finite value ($< 2e^2/h$).
(iii) In the intermediate region of $V_g$,
$G$ has a sharp peak of $2 e^2/h$ in height.
This is due to the coexistence of the coherence between
a dot and lead (Kondo coupling) and that between the dots
(spin-spin coupling). At the peak of $G$,
$\kappa + \widetilde{V}_C = \widetilde{\Delta}$:
$T(\omega)$ has a single peak at $\omega=0$ with the height of unity,
which reflects a coherent transport channel connecting the
left and right leads via the two dots.\cite{Aono01}
Just on the left side of the peak,
$\kappa + \widetilde{V}_C > \widetilde{\Delta}$ and
$T(\omega)$ has double peaks.
On the right side of the peak,
$\kappa + \widetilde{V}_C < \widetilde{\Delta}$.

In Fig.\ \ref{fig:G_MC_J_weak}(b), the MC is shown for three values of
$V_g$ around the peak of $G$.
On the right side of the peak, we observe a negative MC (solid line).
On the left side of the peak,
the MC is positive (broken line).\cite{Kondo_enhance}
They are explained by the above-mentioned peak structures of the
transmission spectrum $T(\omega)$.
At the peak of $G$, the conductance is almost constant at $2e^2/h$
for $g \mu_B B^<_{\sim} T_K/2$ and decreases considerably with $B$ 
for $g \mu_B B^>_{\sim} T_K/2$.
This behavior of MC reflects a flat-topped
single peak structure of $T(\omega)$
with $\kappa + \widetilde{V}_C = \widetilde{\Delta}$.\cite{Aono01}

Finally we look at the case of $V_C / \Delta >1$.
In Fig.\ \ref{fig:MC_strong}, the MC is shown by a dotted line
when $V_C / \Delta = 1.6$ and $J/\Delta = 1.0$.\cite{J}
In this case,
$\kappa + \widetilde{V}_C$ is always larger than $\widetilde{\Delta}$ and
thus $T(\omega)$ has double peaks.
We observe a positive MC as in the case of $J=0$ (solid line).
Note that the Kondo state survives in a larger region of the
magnetic field, in the presence of $J$ than in the absence of $J$,
although $G$ is smaller at small $B$.
This is because the antiferromagnetic spin coupling prevents the
formation of the spin polarized state, and as a result,
keeps the Kondo couplings until larger values of $B$.

In conclusions,
we have investigated the magnetoconductance in
coupled quantum dots in the Kondo region.
The MC illustrates peak structures of the Kondo resonant spectra,
single peak or double peaks, depending on a ratio of $V_C/\Delta$ and
value of $J$.
We expect that the observation of the Kondo resonances by the
MC is easier than that by
the differential conductance under finite source-drain voltages\cite{Aguado99}
because the dephasing processes should influence the Kondo states
in the latter case.\cite{Kaminski}

T.A. acknowledges the support of a fellowship from
Special Postdoctoral Researches Program at RIKEN.
Numerical calculations were performed on the workstation in
the Computer Information Center, RIKEN.


%
%
%
%
%
%


\begin{references}
\bibitem{Tarucha}
S.\ Tarucha {\it et al.},
Phys.\ Rev.\ Lett.\ {\bf 77}, 3613 (1996).

\bibitem{METbook}
{\it Mesoscopic Electron Transport},
edited by  L.\ L.\ Sohn, L.\ P.\ Kouwenhoven and G.\ Sch\"on,
(Kluwer Academic, Dordrecht, 1997).

\bibitem{Oosterkamp} 
T.\ H.\ Oosterkamp {\it et al.},
Nature (London) {\bf 395}, 873 (1998).

\bibitem{Blick}
R.\ H.\ Blick {\it et al.},
Phys.\ Rev.\ Lett.\ {\bf 80}, 4032 (1998);
R.\ H.\ Blick, D.\ W.\ van~der~Weide, R.\ J.\ Haug, and K.\ Eberl,
{\it ibid.} {\bf 81}, 689 (1998).

\bibitem{Goldhaber}
D.\ Goldhaber-Gordon {\it et al.},
Nature (London) {\bf 391}, 156 (1998);
Phy.\ Rev.\ Lett.\ {\bf 81}, 5225 (1998).

\bibitem{Cronenwett}
S.\ M.\ Cronenwett, T.\ H.\ Oosterkamp, and L.\ P.\ Kouwenhoven,
Science {\bf 281}, 540 (1998).


\bibitem{Simmel99}
F.\ Simmel {\it et al.},
Phys.\ Rev.\ Lett.\ {\bf 83}, 804 (1999).

\bibitem{Schmid00}
J.\ Schmid, J.\ Weis, K.\ Eberl, and K.\ v.\ Klitzing,
Phys.\ Rev.\ Lett.\ {\bf 84}, 5824 (2000).

\bibitem{Wilfred00}
W.\ G.\ van~der Wiel {\it et al.},
Science {\bf 289}, 2105 (2000).


\bibitem{Glazman}
 L.\ I.\ Glazman and M.\ {\'E}.\ Ra{\u\i}kh,
Pis'ma Zh.\ Eksp.\ Teor.\ Fiz.\ {\bf 47}, 378 (1998).
[{JETP} Lett.\ {\bf 47}, 452 (1988).]


\bibitem{Ng} T.\ K.\ Ng and P.\ A.\ Lee,
Phys.\ Rev.\ Lett.\ {\bf 61}, 1768 (1988).

\bibitem{Kawabata}
A.\ Kawabata,
J.\ Phys.\ Soc.\ Jpn.\ {\bf 60}, 3222 (1991).

\bibitem{Hershfield}
S.\ Hershfield, J.\ H.\ Davies, and J.\ W.\ Wilkins,
Phys.\ Rev.\ Lett.\ {\bf 67}, 3720 (1991);
Phys.\ Rev.\ B {\bf 46}, 7046 (1992).

\bibitem{Meir93}
Y.\ Meir, N.\ S.\ Wingreen, and P.\ A.\ Lee, Phys.\ Rev.\ Lett.\
{\bf 70}, 2601 (1993).

\bibitem{Wingreen94}
N.\ S.\ Wingreen and Y.\ Meir, Phys.\ Rev.\ B {\bf 49}, 11040 (1994).

\bibitem{Ivanov}
T.\ Ivanov, Europhys.\ Lett.\ {\bf 40}, 183 (1997);
Phys.\ Rev.\ B {\bf 56}, 12339 (1997).

\bibitem{Pohjola}
T.\ Pohjola {\it et al.},
Europhys.\ Lett.\ {\bf 40}, 189 (1997).

\bibitem{Aono98} T.\ Aono, M.\ Eto and K.\ Kawamura,
 J.\ Phys.\ Soc.\ Jpn.\ {\bf 67}, 1860 (1998).

\bibitem{Georges} 
A.\ Georges and Y.\ Meir,
Phys.\ Rev.\ Lett.\ {\bf 82}, 3508 (1999).

\bibitem{Andrei}
N.\ Andrei, G.\ T.\ Zim\'anyi and G.\ Sch\"on,
Phys.\ Rev.\ B {\bf 60}, 5125 (1999).

\bibitem{Busser99}
C.\ A.\ B\"usser {\it et al.},
Phys.\ Rev.\ B {\bf 62}, 9907 (2000).

\bibitem{Aguado99}
R.\ Aguado and D.\ C.\ Langreth,
Phys.\ Rev.\ Lett.\ {\bf 85}, 1946 (2000).


\bibitem{Izumida00}
W.\ Izumida and O.\ Sakai,
Phys.\ Rev.\ B {\bf 62}, 10260 (2000).

\bibitem{Aono01}
T.\ Aono and M.\ Eto,
Phys.\ Rev.\ B {\bf 63}, 125327 (2001).


\bibitem{Jones}
B.\ A.\ Jones, C.\ M.\ Varma, and J.\ W.\ Wilkins,
Phys.\ Rev.\ Lett.\ {\bf 61}, 125 (1988); 
B.\ A.\ Jones and C.\ M.\ Varma, Phys.\ Rev.\ B {\bf 40},
324 (1989).

\bibitem{Jones2}
B.\ A.\ Jones, G.\ Kotliar, and A.\ J.\ Millis,
Phys.\ Rev.\ B {\bf 39}, 3415 (1989).

\bibitem{Sakai90}
O.\ Sakai, Y.\ Shimizu, and T.\ Kasuya,
Solid State Commun.\ {\bf 75}, 81 (1990);
O.\ Sakai and Y.\ Shimizu,
J.\ Phys.\ Soc.\ Jpn.\ {\bf 61}, 2333 (1992);
{\bf 61}, 2348 (1992).


\bibitem{Coleman87}
P.\ Coleman,
 Phys.\ Rev.\ B {\bf 35}, 5072 (1987).

\bibitem{Read83}
N.\ Read and D.\ M.\ Newns,
 J.\ Phys.\ C {\bf 16}, L1055 (1983).


\bibitem{Bickers87}
N.\ E.\ Bickers,
 Rev.\ Mod.\ Phys.\ {\bf 59}, 845 (1987).

\bibitem{Newns88}
D.\ M.\ Newns and N.\ Read,
Adv.\ Phys.\ {\bf 36}, 799 (1988).

\bibitem{HewsonBook}
A.\ C.\ Hewson,
{\it The Kondo Problem to Heavy Fermions}
(Cambridge Univ. Press, Cambridge, 1997).


\bibitem{Kawamura} K.\ Kawamura and T.\ Aono,
 Jpn.\ J.\ Appl.\ Phys.\ {\bf 36}, 3951 (1997).

\bibitem{Kondo_temp}
When $J=0$,
$T_{\rm K} =
(\Delta/\pi) \exp ( \pi E_0 /\Delta) f(V_C /\Delta)$
with $f(x) = \exp \left( x \arctan x \right)
     / \sqrt{ 1 + x^{2}}$.\cite{Georges,Aono01}
The dot level $E_0$ is located below the Fermi level ($E_0<0$).
For larger negative $E_0$, $T_K$ becomes smaller.

\bibitem{J}
We set $J = 4 V_{C}^{2}/U$ where $U/\Delta = 4 \times 10^{2}$ is fixed.

\bibitem{Kondo_enhance}
The positive MC is more pronounced than in Fig.\ \ref{fig:MC_strong}.
In this case, the magnetic field $B$ suppresses the antiferromagnetic
spin coupling $\kappa$ and consequently {\it enhances} the Kondo effect.
With increasing $B$, $\kappa$ decreases and hence
the bonding and anti-bonding peaks of $T(\omega)$ become closer to
each other. This enhances MC besides the Zeeman effect.
At a certain value of $B$, $G= 2 e^2/h$.
As $B$ increases further, $\kappa + \widetilde{V}_C$ becomes
less than $\widetilde{\Delta}$; the molecular levels disappear and
MC is negative.

\bibitem{Kaminski}
A.\ Kaminski, Yu.\ V.\ Nazarov, and L.\ I.\ Glazman,
Phys.\ Rev.\ Lett.\ {\bf 83}, 384 (1999).


\end{references}
\end{document}